\documentclass{ws-procs9x6}

\begin{document}

\title{Magnetized (Shift-)Orientifolds}

\author{Gianfranco Pradisi}

\address{Dipartimento di Fisica, Universita di Roma 
``Tor Vergata''\\
INFN, Sezione di Roma 2\\ 
Via della Ricerca Scientifica 1, 00133 Roma,
Italy\\
E-mail: gianfranco.pradisi@roma2.infn.it}


\maketitle

\vskip 1.5cm

\abstracts{We study four dimensional  
$Z_2 \times Z_2$ (shift)-orientifolds 
in presence of internal magnetic fields and NS-NS 
$B$-field backgrounds, describing in some detail 
one explicit example with N=1 supersymmetry. 
These models are related by $T$-duality to
orientifolds with $D$-branes intersecting at angles 
and exhibit, due to the background fields,  
a rank reduction of the gauge group 
and multiple matter families.
Moreover, the low-energy spectra are chiral and anomaly 
free if $D5$-branes are present along the magnetized directions.}

\vfil
{\hbox{\hskip 8.2cm {\rm ROM2F-02/25}  \hfil}
\hbox{\hskip 8.2cm {\rm hep-th/0210088} \hfil}
\newpage

\section{Introduction}

Within the String/M-theory picture\cite{Witten}, 
the Standard Model of strong and 
electroweak interactions should emerge, together with a consistent 
quantum version of General Relativity, as a low-energy limit of some
vacuum configuration in lower-dimensions.  In other words, there should be
a region in the moduli space of String/M-theory that is connected, 
possibly in a very
complicated way, to our low-energy world.  Unfortunately, we are able to 
control only a small region (actually a zero-measure region) of the 
moduli-space itself, and we miss the dynamical principle that should 
drive us to the choice of the (unique?) ground state.  However, the
investigation of vacua that look as close as possible to the Standard
Model on the one hand can clarify the internal
consistency and help to extract 
some model independent properties of the theory, and on the
other hand can lead 
to some predictions of possible ``experimental signatures'' of the 
underlying String/M-theory structure.  
There has been a lot of effort in these  
directions in the last few years, both in the framework of 
conventional Heterotic SUSY-GUT scenarios\cite{dienes}, 
and in the framework of 
type I models\cite{typei}, or more generally in the context of 
``Brane World'' scenarios\cite{brwor}.

An interesting way to build chiral type I models while
preserving supersymmetry in the (bulk) gravity sector is the
introduction of internal background magnetic fields along some
compactified directions\cite{bachas,aads,ted}.  
A magnetic field along a $U(1)$ subgroup of the
Chan-Paton gauge group affects only the boundary conditions of the
(super)strings in the open sector\cite{ft}, 
providing an energy separation between 
states of different spins.  As a consequence, 
supersymmetry is broken on the branes  
and, generically, some Nielsen-Olesen instabilities manifest 
themselves by the appearance of tachyonic excitations.  This
kind of deformation is connected by (open) T-duality to type I vacua
with $D$-branes intersecting at angles\cite{douglas}, and thus no longer 
parallel to the corresponding O-planes, and  
seems to be the most promising proposal to recover 
(some extension of) the Standard Model in the low-energy limit
\cite{ted,csu,phen1,phen2,phen3,phen4,phen5,phen6}.  
The introduction of a pair of aligned internal magnetic fields allows 
configurations with non-vanishing instanton
density in the compact internal space, 
namely self-dual or antiself-dual field strenghts\cite{aads}. 
The resulting
$D9$-branes contributions to the 
tension and $R-R$ charge can be exactly 
compensated by the introduction of suitable lower dimensional 
branes, eliminating the corresponding tachyonic instabilities and 
recovering supersymmetry if the BPS bound is restored, or giving rise
to non-BPS configurations 
(models with ``brane supersymmetry breaking''\cite{bsb,aadds,su})
if the fields carry $R-R$ charges that mimic
anti-$D$-branes.  In the T-dual  picture, they correspond to vacua with
branes intersecting at
very special angles, the ones exactly preserving supersymmetry or their
opposites.  
A chiral spectrum is obtained if there are massless 
open-string states in bifundamental representations 
identified by the magnetic field and the inequivalent
$D5$-branes or, in other words, at the intersection of 
magnetized $D9$-branes and those $D5$-branes whose world-volume 
invades at least one of the magnetized tori.

In this talk we shall describe a class
of four-dimensional freely-acting orientifolds,  using as guiding 
example a chiral four-dimensional supersymmetric model.  
These models are obtained deforming the $Z_2 \times Z_2$ shift-orientifolds 
described in ref.\cite{freely} with internal (open) background magnetic
fields and will be
discussed in more detail in ref.\cite{our}.  
Their closed oriented massless spectra 
are reported in Table 
\ref{tab1} and, as can be observed analyzing the last column, 
they correspond 
to compactifications of type IIB superstrings on (singular limits of) 
Calabi-Yau manifolds.
\begin{table}[th]
\tbl
{Closed oriented spectra of $Z_2 \times Z_2$ shift-orbifolds.
\vspace*{1pt}}
{\footnotesize
\begin{tabular}{|c||c|c|c|c|c|c|}
\hline
{}&{\rm untwisted}&{\rm untwisted}&
{untwisted}&{twisted}&{twisted}&{}\\ 
{\rm model}&{\rm SUGRA}&{H}&{V}&{H}&{V}&{}\\
\hline
{$p_3$}&{$N=2$}&{$1+3$}&{3}&{$16$}&{$16$}&{CY $(19,19)$}\\\hline\hline
{$p_2 p_3$}&{$N=2$}&{$1+3$}&{3}&{$8$}&{$8$}&{CY $(11,11)$}\\\hline
{$w_2 p_3$}&{$N=2$}&{$1+3$}&{3}&{$8$}&{$8$}&{CY $(11,11)$}\\\hline
{$w_1 p_2$}&{$N=2$}&{$1+3$}&{3}&{$8$}&{$8$}&{CY $(11,11)$}\\\hline\hline
{$p_1 p_2 p_3$}&{$N=2$}&{$1+3$}&{3}&{$0$}&{$0$}&{CY $(3,3)$}\\\hline
{$p_1 w_2 w_3$}&{$N=2$}&{$1+3$}&{3}&{$0$}&{$0$}&{CY $(3,3)$}\\\hline
{$w_1 p_2 p_3$}&{$N=2$}&{$1+3$}&{3}&{$0$}&{$0$}&{CY $(3,3)$}\\\hline
{$w_1 p_2 w_3$}&{$N=2$}&{$1+3$}&{3}&{$0$}&{$0$}&{CY $(3,3)$}\\\hline
{$w_1 w_2 p_3$}&{$N=2$}&{$1+3$}&{3}&{$0$}&{$0$}&{CY $(3,3)$}\\\hline
{$w_1 w_2 w_3$}&{$N=2$}&{$1+3$}&{3}&{$0$}&{$0$}&{CY $(3,3)$}\\\hline
\end{tabular}\label{tab1} }
\end{table}
As shown in ref.\cite{freely}, the resulting orientifolds give rise to 
a rich but non-chiral 
class of models with partial breaking of supersymmetry and
interesting brane configurations.  Chirality can be obtained,
in some cases, introducing magnetic deformations, as we are going to 
discuss in the next section.
\section{The $w_2 p_3$ models}
In what follows we shall consider the $w_2 p_3$ 
model, that captures all the basic 
features of this class of orientifolds.  The starting point is an orbifold of 
the type IIB superstring 
compactified on a six-torus taken, in a self-explanatory 
notation, as the product $T^{45} \times T^{67} \times T^{89}$, with complex
coordinates $(Z_{1},Z_{2},Z_{3})$, where each
two-torus can be equipped with a $NS-NS$ two-form $B_{i}$ of rank $r_i$ 
(with $r_i = 0$ or $2$), 
quantized\cite{toroidal,carlo} 
if the orientifold projection is to be performed.  The 
orbifold group is the combination of the $Z_2 \times Z_2$ generated by
$g=(+,-,-)$ and $h=(-,-,+)$, where the minus signs correspond to
a conventional two-dimensional $Z_2$ inversion 
$(Z_i\rightarrow -Z_i)$, with a winding shift 
along the $6$-th (real) direction 
and a momentum shift along the $8$-th (real) direction.
As a consequence, there are no fixed points while the 
amplitudes corresponding to the discrete
torsion orbit of the modular group are absent.  
The unoriented projection of the closed spectrum in Table \ref{tab1}, 
obtained by the action of the world-sheet parity 
operator $\Omega$, produces a Klein bottle amplitude 
containing $O9_+$, $O5_{1,+}$ and $O5_{2,+}$, since 
the shifts lift the massless states along 
the $T^{89}$ direction, thus eliminating the corresponding 
$O5_{3,+}$ present in the ``plain'' $Z_2 \times Z_2$ open 
descendants\cite{bianchi,aadds,freely}.  
The resulting truncation of the closed spectrum is 
displayed in Table \ref{tab2}.
\begin{table}[th]
\tbl
{Closed unoriented spectra of $w_2 p_3$ models.
\vspace*{1pt}}
{\footnotesize
\begin{tabular}{|c||c|c|c|c|}
\hline
{\rm $B$ rank}&{\rm untwisted}&{untwisted}&{twisted}&{twisted}\\ 
{\rm $r_2+r_3$}&{\rm SUGRA}&{C}&{C}&{V}\\\hline\hline
{$0$}&{$N=1$}&{$1+3+3$}&{$8+8$}&{$0$}\\\hline
{$2$}&{$N=1$}&{$1+3+3$}&{$6+6$}&{$2+2$}\\\hline
{$4$}&{$N=1$}&{$1+3+3$}&{$5+5$}&{$3+3$}\\\hline
\end{tabular}\label{tab2} }
\end{table}
Notice that the low energy effective 
$N=1$ supergravities are coupled to 
different numbers of chiral and vector multiplets from twisted
sectors, depending on the rank of the $NS-NS$ two-form blocks along the
directions affected by the shifts.  In order to 
neutralize the net $R-R$ charge 
of the background, the annulus amplitude in the presence of a pair of 
$U(1)$ (aligned) magnetic fields along $Z_{2}$ and $Z_{3}$ directions 
and of the 
discretized $B$-field must contain the ubiquitous $D9$-branes, one set of
``magnetized'' $D9$-branes ($D9_m$-branes) 
and two sets of $D5$-branes.  These are the $D5_1$-branes, whose 
world-volume invades the internal $Z_{1}$ coordinates, and the 
$D5_2$-branes, whose 
world-volume invades the internal $Z_{2}$ coordinates, as requested by the
presence of the corresponding O-planes.  Indicating by 
$(n,m,d_1,d_2)$ the Chan-Paton charge multiplicities, 
or equivalently the numbers of
$D9$-branes, $D9_m$-branes, $D5_1$-branes and $D5_2$-branes,
there are several solutions due to the presence of $B$.  In particular,
the Chan-Paton gauge group can be chosen as in Table \ref{tab3}, and  
we shall limit 
the discussion to the complex-charges case a). 
The interested reader can find a more detailed description 
in the forthcoming ref.\cite{our}.
\begin{table}[th]
\tbl
{Chan-Paton group: a) complex charges; b) real charges.
\vspace*{1pt}}
{\footnotesize
\begin{tabular}{cccccccc}
\hline
{a)}&
{$U(n)$}&
{$\otimes$}&
{$U(d_1)$}&
{$\otimes$}&
{$\bigl(\begin{array}{c} USp(d_2) \\ SO(d_2) \end{array} \bigr)$}&
{$\otimes$}&
{$U(m)$}\\\hline\hline
{b)}&
{$\bigl(\begin{array}{c} USp(n_1) \times Usp(n_2) \\ 
SO(n_1) \times SO(n_2) \end{array} \bigr)$}&
{$\otimes$}&
{$\bigl(\begin{array}{c} USp(d_1) \times  USp(d_2) \\ 
SO(d_1)  \times SO(d_2) \end{array} \bigr)$}&
{$\otimes$}&
{$\bigl(\begin{array}{c} USp(d_3) \\ SO(d_3) \end{array} \bigr)$}&
{$\otimes$}&
{$U(m)$}\\\hline
\end{tabular}\label{tab3} }
\end{table}
For generic values of the magnetic fields, supersymmetry is broken and
Nielsen-Olesen instabilities appear in the form of open tachyonic 
excitations. However, for self-dual configurations of the field or, in
a T-dual language, for special choices of the angles between the 
$D6$-branes, one supersymmetry is still preserved and tachyons are 
absent.  In this case, the tadpole cancellation conditions read
\begin{eqnarray}
n +  \bar{n} + m  + \bar{m}  & = & 16  \ 2^{-r/2} ,     \nonumber \\
d_1 + \bar{d_1} + 2^{r_2/2 + r_3/2} \ | k_{2} k_{3} | 
\ (m + \bar{m})  & = & 16 \  2^{-r/2} , \nonumber \\
d_2 & = & 8  \ 2^{-r/2} ,  
\label{tadw2p3}
\end{eqnarray}
with $n =  \bar{n}$, $m=\bar{m}$ and $d_1 = \bar{d_1}$, where 
$r=r_1+r_2+r_3$ is the total rank of $B$ 
and $k_{2}$ and $k_{3}$ are the Landau-level degeneracies along the
corresponding directions.  They count the numbers of zero-modes of the
magnetized open strings with non-vanishing total magnetic charge.  It should 
be stressed that the $m$-charges contribute to the tadpoles of the $R-R$ 
ten-form and the $R-R$ six-form.  This signals the phenomenon of 
``brane transmutation''\cite{aads}, 
connected to the Wess-Zumino-like term in the
$D$-brane action.  According to it, a magnetized $D9$-brane with 
non-vanishing instanton number mimics the
behaviour of a stack of $k_{2} k_{3}$ $D5$-branes, thus producing
both a rank reduction of the Chan-Paton group and the
presence of multiple families of matter fields.  
It can also be interpreted\cite{aads,csu} as the inverse 
small instanton 
transition:  the stack of $D5$-branes dissolves into a $D9$-brane that, being
a ``fat'' instanton, invades the whole internal 
space and gives rise to a transition
along some flat directions.  In the T-dual language, it
corresponds to a recombination of $D6$-branes wrapping different intersecting 
cycles.
The open unoriented spectra are reported in Table \ref{tab4}, 
\begin{table}[th]
\tbl
{Open spectra of $w_2 p_3$ models (complex charges).
\vspace*{1pt}}
{\footnotesize
\begin{tabular}{|c|c|c|}
\hline
{\rm Mult.}&{Number}&{\rm Reps.}\\\hline
{\rm $C$}&{$1$}&{$(Adj,1,1,1),(1,Adj,1,1)$}\\
{}&{}&{$(1,1,1,Adj)$}\\\hline
{\rm $C$}&{$\biggl(\begin{array}{c}2\\0\end{array} \biggr)$ or 
$\biggl(\begin{array}{c}0\\2\end{array} \biggr)$}&
{$\begin{array}{c} (A+\bar{A},1,1,1)\\(S+\bar{S},1,1,1) \end{array}$,
$\begin{array}{c} (1,A+\bar{A},1,1)\\ (1,S+ \bar{S},1,1) \end{array}$}\\\hline
{\rm $C$}&{$3$}&{$(1,1,A,1)$ or $(1,1,S,1)$}\\\hline
{\rm $C$}&{$2^{r_2+r_3} \ |k_2 \, k_3| + 2$}&{$(F,1,1,F)$,
$(\bar{F},1,1,\bar{F})$}\\\hline
{\rm $C$}&{$2^{r_2+r_3} \ |k_2 \, k_3| - 2$}&{$(\bar{F},1,1,F)$,
$(F,1,1,\bar{F})$}\\\hline
{\rm $C$}&{$2 \ 2^{\frac{r_2+r_3}{2}}$}&{$(F,F,1,1)$,$(\bar{F},
\bar{F},1,1)$}\\\hline
{\rm $C$}&{$2 \ 2^{\frac{r_2+r_3}{2}}$}&{$(1,\bar{F},1,F)$,
$(1,F,1\bar{F})$}\\\hline
{\rm $C$}&{$2^{r_2+r_3} \ 2 \ |k_2 \, k_3| + 1 + 2^{\frac{r_2+r_3}{2}}
\eta_1 |k_2 \, k_3| + \eta_1 + 2^{\frac{r_2}{2}} |k_2|$}
&{$(1,1,1,A)$}\\\hline
{\rm $C$}&{$2^{r_2+r_3} \ 2 \ |k_2 \, k_3| + 1 - 2^{\frac{r_2+r_3}{2}}
\eta_1 |k_2 \, k_3| - \eta_1 - 2^{\frac{r_2}{2}} |k_2|$}
&{$(1,1,1,S)$}\\\hline
{\rm $C$}&{$2^{r_2+r_3} \ 2 \ |k_2 \, k_3| + 1 + 2^{\frac{r_2+r_3}{2}}
\eta_1 |k_2 \, k_3| + \eta_1 - 2^{\frac{r_2}{2}} |k_2|$}
&{$(1,1,1,\bar{A})$}\\\hline
{\rm $C$}&{$2^{r_2+r_3} \ 2 \ |k_2 \, k_3| + 1 - 2^{\frac{r_2+r_3}{2}}
\eta_1 |k_2 \, k_3| - \eta_1 + 2^{\frac{r_2}{2}} |k_2|$}
&{$(1,1,1,\bar{S})$}\\\hline\hline
{\rm $C_L$}&{$2^{\frac{r_1+r_3}{2}+r_2} \ 2 \ |k_2|$}
&{$(1,1,F,F)$}\\\hline
\end{tabular}\label{tab4} }
\end{table}
where $\eta_1$ is a free sign and $A$, $S$ and $F$ 
stand for the antisymmetric, the symmetric and the fundamental 
representations of the gauge group in Table \ref{tab3}, respectively.
Notice that, in the presence of $D5_2$ branes, one gets an extra-bonus, due
again to the magnetic field: chirality.  
Chiral fermions lie at the intersection of $D$-branes and, as evident
from Table \ref{tab4}, the chiral sector is exactly the one related to open
strings of the $(d_2,m)$ type thus coming from the intersection between
magnetized $D9$-branes and $D5_2$-branes extended along one of the 
directions affected by the magnetic field.  
Of course, one could add brane-antibrane pairs and
suitable combinations of Wilson lines in order to produce string vacua that
exhibit low-energy spectra 
as close as possible to the Standard Model, both for the field content 
and for the phases of the various symmetries
\cite{ted,csu,phen1,phen2,phen3,phen4,phen5,phen6}.  
The dynamical stability of all these vacua
is still an open problem, due to the presence of
$NS-NS$ tadpoles when supersymmetry is broken.  The 
principle (if any) according to which Nature selects the right vacuum is
still lacking, and its quest is probably the most important challenge
in the String/M-theory research activity.
\section*{Acknowledgments}
I would like to thank the Organizers of the ``The First International 
Conference On String Phenomenology'' 
for the kind invitation.  It is a pleasure to thank 
M. Larosa for the enjoyable collaboration, A. Sagnotti for the 
many discussions and collaboration at early stages of this 
research and 
C. Angelantonj, M. Bianchi, R. Blumenhagen, G. D'Appollonio, 
E. Dudas, J. Mourad and Ya.S. Stanev for illuminating discussions.  
This work was supported in part by I.N.F.N., by the 
European Commission RTN programmes HPRN-CT-2000-00122 and 
HPRN-CT-2000-00148, by the INTAS contract 
99-1-590, by the MURST-COFIN contract 2001-025492 and by the NATO
contract PST.CLG.978785.

\end{document}